\documentclass[structabstract]{aa}  
\usepackage{graphicx}
\usepackage{txfonts}
\usepackage{natbib}

\newcommand{\fig}[1]{Fig.~\ref{#1}}
\newcommand{\sect}[1]{Sect.~\ref{#1}}
\def\spose#1{\hbox to 0pt{#1\hss}}
\def\gsim{\mathrel{\spose{\lower 3pt\hbox{$\mathchar"218$}}
          \raise 2.0pt\hbox{$\mathchar"13E$}}}
\def\lsim{\mathrel{\spose{\lower 3pt\hbox{$\mathchar"218$}}
          \raise 2.0pt\hbox{$\mathchar"13C$}}}

\begin{document}

\title{The origin of the late rebrightening in GRB 080503}

\author{
   	R. Hasco\"et, 
         F. Daigne \thanks{Institut Universitaire de France}
          \and
          R. Mochkovitch
          }

\institute{UPMC-CNRS, UMR7095, Institut d'Astrophysique de Paris, F-75014, Paris, France \\
               \email{[hascoet;daigne;mochko]@iap.fr}}

\abstract
{GRB 080503, detected by \textit{Swift}, belongs to the class of bursts whose prompt phase consists of an initial short spike followed by 
a longer soft tail. It did not show any transition to a regular afterglow at the end of the 
prompt emission but exhibited a surprising rebrightening after one day.}         
{We aim to explain this rebrightening with two different scenarios -- refreshed shocks or a density clump in the circumburst medium --
and two models for the origin of the afterglow, the standard one where it comes from the forward shock, and an
alternative one where it results from a long-lived reverse shock.}
{We computed afterglow light curves either using a single-zone approximation for the shocked region or a detailed multi-zone method that more accurately accounts for
the compression of the material.}
{We find that in several of the considered cases the detailed
model must be used to obtain a reliable description of the shock dynamics. The density clump scenario is not favored. We confirm previous results that the presence of the clump 
has little effect on the forward shock emission, except if the microphysics parameters evolve when the shock 
enters the clump. Moreover, we find that the rebrightening from the reverse shock is also too weak when it is calculated 
with the multi-zone method. On the other hand, in the refreshed-shock scenario both the forward and reverse shock models provide satisfactory fits of 
the data under some additional conditions on the distribution of the Lorentz factor in the ejecta and
the beaming angle of the relativistic outflow.}
{}

\keywords{Gamma rays bursts: general; Gamma rays bursts: individual: GRB 080503; 
Shock waves; Radiation mechanisms: non thermal}
 
\authorrunning{
R. Hasco\"et~et~al. 
          } 
\titlerunning{The origin of the late rebrightening in GRB 080503}

\maketitle
\section{Introduction}
Short bursts with a duration of less than 2 s represent about 25\% of the BATSE sample \citep{kouveliotou:1993} but
had to wait until 2005 (i.e. eight years after long bursts) to enter the afterglow era \citep{gehrels:2005, fox:2005, hjorth:2005, berger:2005}. 
This is due to two reasons: (\textit{i}) short bursts tend to emit less photons because of harder spectra and lower fluences \citep{kouveliotou:1993}, which makes their localization more difficult; (\textit{ii}) they have fainter afterglows, which are harder to detect. Following the discovery of the first afterglows, it appeared that the nature of the host galaxy, the location of the afterglow, and the absence of a supernova imprint in the visible light curve (even when the host is located at a redshift below 0.5) were indicative of progenitors that were different from those of 
long bursts \citep{gehrels:2005, fox:2005, soderberg:2006}. Several short burts are clearly associated to elliptical galaxies \citep{bloom:2006, berger:2009} while others with accurate
positions appear to have no coincident hosts, which clearly excludes progenitors belonging to the young
population and favors merger scenarios involving compact objects \citep{narayan:1992, mochkovitch:1993,ruffert:1999, belczynski:2006}.

About 40\% of the short bursts have no detectable afterglows after about 1000 s while the other 60\% \citep{sakamoto:2009} have 
long-lasting afterglows comparable to those of long bursts (see the review on short bursts by \citealt{nakar:2007b} and references therein). If short bursts indeed result from the merging of two 
compact objects, the kick received when the black hole or neutron star components formed
in past supernova explosions \citep{hobbs:2005, belczynski:2006} can allow the system to reach the low-density outskirts of the host galaxy (or even
to leave the galaxy) before coalescence occurs. This can naturally explains why 
some afterglows are so dim or have no coincident host (the observational data presented in \citealt{troja:2008} show that the galactocentric offset of short bursts is on average much larger than for long bursts).

The direct and simple connection between duration and progenitor class became fuzzier
when it was found that in some bursts an initial short duration spike is followed by
a soft tail lasting several tens of seconds \citep{barthelmy:2005, villasenor:2005, norris:2006}. It was then suggested \citep{zhang:2006, gehrels:2006} to introduce a new terminology 
that would distinguish type-I bursts resulting from mergers and type-II events coming from collapsars. In the absence of a detected afterglow that can help to relate the burst 
to either the old or young stellar population, a vanishing spectral lag (for both the spike 
and the extended emission) has been proposed as an
indicator for a type-I identification \citep{gehrels:2006}.  

GRB 080503 belongs to the class of short bursts with extended emission. The extended emission 
ended with a steep decay that was not immediately 
followed by a standard afterglow component. A peculiar feature in GRB 080503 is that after 
remaining undetected for about one day, it showed 
a spectacular rebrightening (both in X-rays and the visible), which could be followed for five days in the visible. \citet{perley:2009}
described in great detail the multi-wavelength data they collected for this event and discussed different possibilities
that could account for the late rebrightening: ({\it i}) a delayed rise of the afterglow due to an extremely low density
of the surrounding medium; ({\it ii}) the presence of a density clump in the burst environment; 
({\it iii}) an off-axis jet that becomes visible 
when relativistic beaming has been reduced by deceleration (see e.g. \citealt{granot:2002}); ({\it iv}) a refreshed shock, when a slower part of the ejecta 
catches up with the shock, again as a result of deceleration (see e.g. \citealt{sari:2000}); and finally ({\it v}) a ``mini-supernova''  
from a small amount of ejected material powered by the decay of $^{56}$Ni \citep{li:1998}. 

Case ({\it i}) imposes an external density below $10^{-6}$ cm$^{-3}$, which seems unreasonably low; case ({\it iii}) implies
a double-jet structure \citep{granot:2005} with one on-axis component producing the prompt emission but no visible afterglow (which can be possible
only if the prompt phase has a very high efficiency) and the other one (off-axis) producing the delayed 
afterglow; case ({\it v}) can account for the rebrightening in the visible, but not in X-rays.
   
We therefore reconsider in this work  the two most promising cases ({\it ii}) and ({\it iv}) in the context of the standard model where the afterglow is produced by
the forward shock \citep{meszaros:1997,sari:1998} but also in the alternative one where it comes from a long-lived reverse shock \citep{genet:2007,uhm:2007}.
The paper is organized as follows. We briefly summarize the observational data in \sect{sect_obs_data} and list in \sect{sect_two_comp} 
possible sources for the initial spike and extended emission.
We constrain in \sect{sect_kinen_outflow} the energy released in these two components and discuss in \sect{sect_modeling} 
different ways to explain the rebrightening with a special emphasis on cases ({\it ii}) and ({\it iv}) above. 
Finally \sect{sect_conclusions} is our conclusion.

\section{Short summary of the observational data}
\label{sect_obs_data}

\subsection{Prompt emission}
The {\it Swift}-BAT light-curve of GRB 080503 consists of a short bright initial spike
followed by a soft extended emission of 
respective durations $t_{90,{\rm spike}} =0.32 \pm 0.07$ s and $t_{90,{\rm ee}} =170 \pm 40$ s
\citep{mao:2008, perley:2009}.
The fluence of the extended emission from 5 to 140 s and between 15 and 150 keV was
$S_{\rm ee}^{15 - 150}=(1.86\pm 0.14)\,10^{-6}$ erg.cm$^{-2}$ while the fluence of the
spike $S_{\rm spike}^{15 - 150}$ was 30 times lower. The spectra of both
the spike and the extended  emission were fitted by single power-laws
with respective photon indices $1.59\pm 0.28$ and $1.91\pm 0.12$. The position of the
initial spike in the duration-hardness diagram and the absence of any significant spectral lag 
(together with the absence of a candidate host galaxy directly at the burst
location) make it consistent with a short (type I) burst classification, resulting from the
merging of two compact objects. No spectral lag ana\-lysis could be performed on the
extended emission, which was weaker and softer than the spike. 
\subsection{Afterglow emission}
The afterglow of GRB 080503 was very peculiar. The prompt extended emission ended in X-rays
with a steep decay phase of temporal index $\alpha=2 - 4$ ($F(t)\propto t^{-\alpha}$),
which is common to most long and short bursts. This decay did not show any transition 
to a ``regular afterglow'' and went below the detection limit in less than one hour. This behavior has 
been observed in about 40\% of the short burst population \citep{sakamoto:2009} but in GRB 080503 it covered nearly
six orders of magnitude. In the visible, except for a single Gemini $\it {g}$ band detection at 0.05 day, the
afterglow remained undetected until it exhibited a surprising late rebrightening (both in X-rays 
and the visible) starting at about one day after trigger. Following the peak of the rebrightening,
the available optical data points (extending up to five days) and subsequent upper limits show a steep 
decay of temporal index $\alpha \sim 2$ \citep{perley:2009}.

\section{Origin of the different emission components}
\label{sect_two_comp}

The different temporal and spectral properties of the prompt initial spike and
extended emission indicate that they are produced by distinct parts of the 
outflow, possibly even with different dissipation or radiative mechanisms. 
The temporal structure of the extended emission, showing a short time-scale
variability (with $t_{\rm var}\lsim$ 1 s), excludes the possibility of any 
conventional afterglow origin. Models of the central engine have been proposed, which are able to 
produce a relativistic outflow made of two distinct components 
with kinetic powers
and temporal properties similar to what is seen in short GRBs with extended emission.  
For example, in compact binary progenitors, the extended 
emission could be caused by the fallback of material, following coalescence \citep{rosswog:2007, troja:2008}.
For a magnetar progenitor, \citet{metzger:2008} suggested that the initial spike is produced by accretion
onto the protomagnetar from a small disk, while the extended emission comes from rotational energy extracted 
on a longer time scale. Finally, \citet{barkov:2011} recently described a two-component jet model
that could explain short GRBs both with and without extended emission, where a wide, short-lived jet is powered by
$\nu{\bar \nu}$ annihilation and a narrow, long-lived one by the Blandford-Znajek mechanism.

For the rest of this study it will be assumed that the outflow in GRB 080503 consisted of two main sub-components,
 responsible for the initial spike and the extended emission, respectively, and that the afterglow emission is associated to the interaction of this structured outflow with the circumburst medium. The energy content of each component can be
estimated from the observed fluences (\sect{sect_kinen_outflow}). For the refreshed-shock scenario (see \sect{refreshed_shocks} below) their typical Lorentz factors 
are somewhat constrained by the time of the rebrightening for a given value of the external density. 

\section{Kinetic energy of the outflow}
\label{sect_kinen_outflow}

To obtain the kinetic energy carried by the different parts of the outflow, 
one should start estimating the
correction factor between the 15 - 150 keV and bolometric fluences $C^{\rm bol} = S^{\rm bol} / S^{15-150}$ for both components.
Unfortunately, the shape of the spectrum is poorly constrained so that we will simply assume 
that $ 2 < C^{\rm bol} < 4$. This is the range obtained with the simplifying assumption that the spectrum can be represented 
by a broken power-law of low  and high-energy photon indices $\alpha=-1.5$, $\beta=-2.5$, and peak energy between
20 and 300 keV (with $C^{\rm {bol}
}\sim 2$ - 2.5 for $E_{\rm p}$ between 20 and 100 keV and rising to 4 at $E_{\rm p}=300$ keV).
From the fluence, we can express the total isotropic energy release in gamma-rays as a function of redshift
\begin{equation}
{\cal E}_{\gamma, {\rm iso}}={4\pi D_{\rm L}^2(z)S^{\rm bol}\over 1+z} \, ,
\end{equation}
where $D_{\rm L}(z)$ is the luminosity distance. To finally obtain the kinetic energy, one has to
assume a radiative efficiency $f_{\rm rad}$, defined as the fraction of the initial kinetic 
energy of the flow eventually converted to gamma-rays. The remaining energy at the end of the prompt phase
is then given by
\begin{equation}
{\cal E}_{{\rm K}, {\rm iso}}={1-f_{\rm rad}\over f_{\rm rad}}\,{\cal E}_{\gamma, {\rm iso}}\ .
\end{equation}  
We adopted $f_{\rm rad}\approx 0.1$ as a typical value. It could be lower for internal shocks
\citep{rees:1994, daigne:1998} or higher for magnetic reconnection 
\citep{spruit:2001, drenkhahn:2002, giannios:2006, mckinney:2010} or modified photospheric emission
\citep{rees:2005, beloborodov:2010}. We did not consider scenarii where the radiative efficiency
would be very different for the spike and extended emission even if this possibility cannot be excluded a priori.
Because the redshift of GRB 080503 is not known, we adopted $z=0.5$ as a ``typical'' value for a type-I burst for  the different examples considered in \sect{sect_modeling}. 
This yields ${\cal E}_{{\rm K}, {\rm iso}} \simeq C^{\rm bol}_{\rm ee} \times 1.1\,10^{52} \ \mathrm{erg}$ 
and $C^{\rm bol}_{\rm spike} \times 3\,10^{50} \ \mathrm{erg}$ 
for the extended emission and spike. 
The dominant uncertainties on these energies clearly 
come from the unknown radiative efficiency and distance of the burst. 
We briefly discuss below how our results are affected when assuming a different redshift or a different radiative efficiency.

\section{Modeling the afterglow of GRB 080503}
\label{sect_modeling}

\subsection{Forward and long-lived reverse shocks}
We considered two different mechanisms that can explain GRB afterglows.
The first one corresponds to the standard picture where the afterglow results from
the forward shock propagating in the external medium, following the initial energy deposition  
by the central engine \citep{sari:1998}.
The second one was proposed by \citet{genet:2007} and \citet{uhm:2007} to
account for some of the unexpected features revealed by {\it Swift} observations of the early afterglow.
It consi\-ders that the forward shock is still present but radiatively inefficient and that the emission 
comes from the reverse shock that sweeps back into the ejecta as it is decelerated. 
The reverse shock is long-lived because it is supposed that the ejecta contains a tail of material with low Lorentz factor
(possibly going down to $\Gamma=1$). 
 
We performed the afterglow simulations  
using two different methods to model the shocked material. In the first one it is represented 
by one single zone as in \citet{sari:1998}: the physical conditions just behind the shock 
are applied to the whole shocked 
material. At any given time, all shocked electrons are considered as a single population, injected at the shock
with a power-law energy distribution. Then the corresponding synchrotron spectrum can be calculated, 
taking into account the effect of electron cooling over a dynamical timescale. 
The second method is more accurate, considering separately the evolution of each elementary shocked shell 
\citep{beloborodov:2005} except for the pressure, which is uniform throughout the whole 
shocked ejecta. The electron population (power-law distribution) and magnetic field of each newly shocked shell are computed taking into account the corresponding shock physical conditions and microphysics parameters. Then each electron population is followed individually 
during the whole evolution, starting from the moment of injection, and taking into account radiative 
and adiabatic cooling. The evolution of the magnetic field -- assuming that the toroidal component is dominant -- is estimated 
using the flux conservation condition. Furthermore, it was checked that the magnetic energy density never 
exceeds equipartition.
  
Finally, we made a few more assumptions to somewhat
restrict the parameter space of the study. We adopted a uniform external medium of low density because
GRB 080503 was probably a type-I burst, resulting from the coalescence of two compact objects in a binary
system at the
periphery of its host galaxy. We also assumed that the redistribution microphysics parameters $\epsilon_e$ and 
$\epsilon_B$ 
-- respectively the fraction of the shock dissipated energy that is injected in the population of accelerated relativistic electrons (power-law distribution with a slope $-p$) and in the amplified magnetic field -- 
follow the prescription $\epsilon_e=\epsilon_B^{1/2}$, which results 
from the acceleration process of electrons moving toward current filaments in the
shocked material \citep{medvedev:2006}. This assumption simplifies the discussion but is not critical for 
the general conclusions of our study.

We did not try to fit the initial steep decay in X-rays because it is generally interpreted as the high-latitude
emission ending the prompt phase and not as a true afterglow component. In that respect, it is not clear
if the optical data point at $\sim 0.05$ day should be associated to the high-latitude emission or already
belongs to the afterglow. We assumed that it is of afterglow origin (the most constraining option) 
and imposed that the simulated light curve goes through it. This leads to some specific consequences, mainly
for the reverse shock model (see discussion in \sect{sect_refreshed_rs}).   
\subsection{Refreshed shocks}
\label{refreshed_shocks}

One way to explain the late rebrightening is to consider that the 
forward or reverse shocks have been refreshed by a late supply of energy \citep{rees:1998, 
sari:2000}. This is possible if the initial short duration spike in the burst profile
was produced by a ``fast'' relativistic outflow (of Lorentz factor $\Gamma_{\rm spike}$) while the extended emission
came from ``slower'' material with $\Gamma_{\rm ee}<\Gamma_{\rm spike}$. Then, at early times, only the fast part
of the flow is decelerated and contributes to the afterglow. When the slower part is finally able to catch up, 
energy is added to the shocks and the emission is rebrightened. 

\subsubsection{Forward shock model}
\label{sect_refreshed_fs}

In the standard forward shock model the lack of any detectable afterglow component before one day imposes severe
constraints on either the density of the external medium or the values of the microphysics parameters. Fixing 
$\epsilon_e$ and $\epsilon_B$ to the commonly used values 0.1 and 0.01 implies to take $n\lsim 10^{-6}$ cm$^{-3}$
\citep{perley:2009}. This very low density would likely correspond to the intergalactic medium, which might
be consistent with the absence of a candidate host galaxy down to a visual magnitude of 
$28.5$. We preferred to adopt a less
extreme value $n=10^{-3}$ cm$^{-3}$, more typical of the interstellar medium at the outskirts of a galaxy (see e.g. \citealt{steidel:2010}).
Then, decreasing the microphysics parameters to $\epsilon_e=\epsilon_B^{1/2}=0.05$ becomes necessary to  
remain consistent with the data.

\begin{figure*}
\begin{tabular}{cc}
\includegraphics[width=0.45\textwidth]{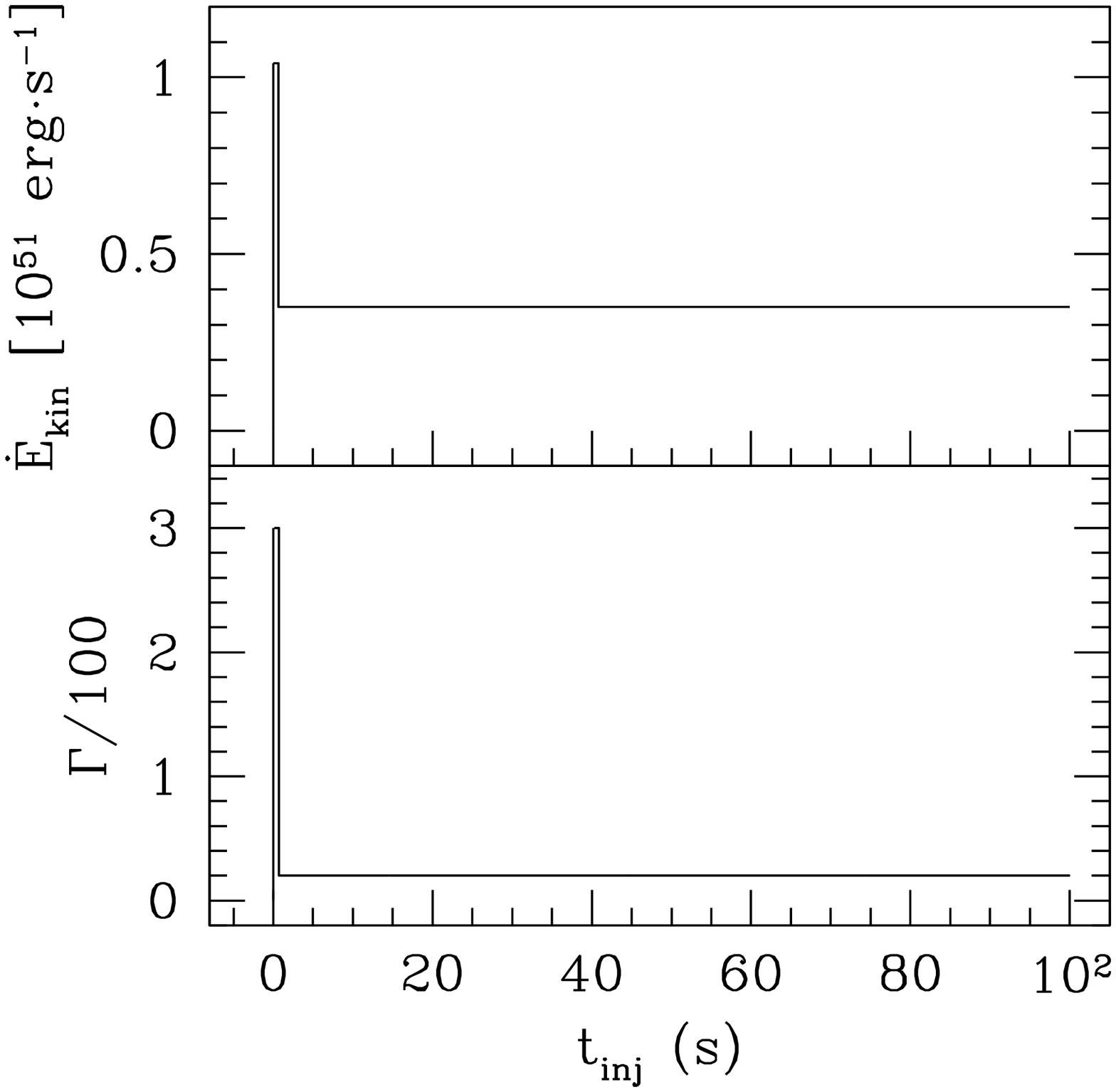} & \includegraphics[width=0.45\textwidth]{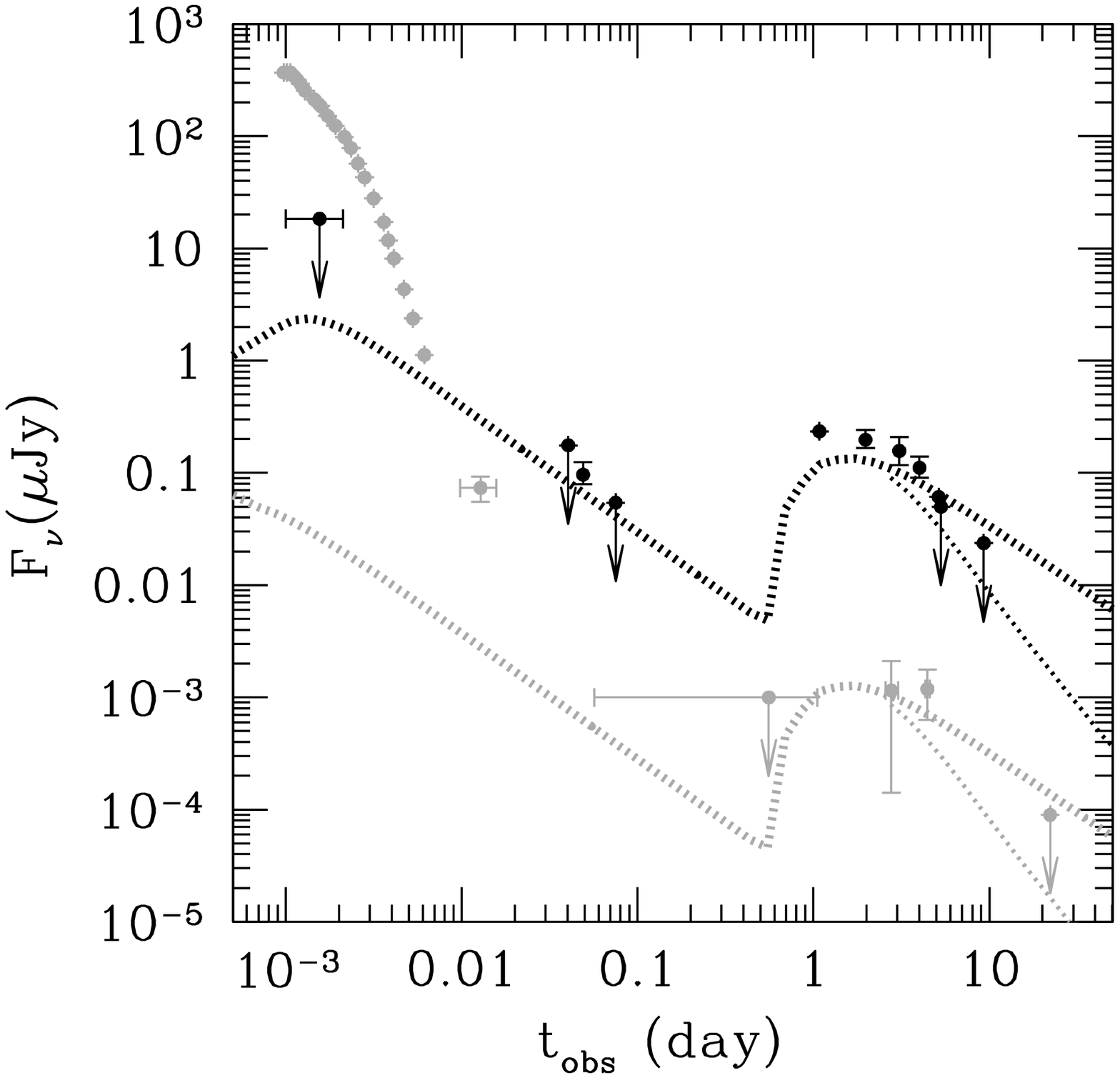} 
\end{tabular}
\caption{\textbf{Refreshed shocks: forward shock model.} 
\textit{Left panel}: 
initial distribution of the Lorentz factor (lower part) and kinetic power (upper part) in the flow as a function 
of injection time $t_{\rm inj}$. 
\textit{Right panel}: 
synthetic light curves at 2 eV (black, dotted line) and 10 keV 
(gray, dotted line) compared to the data from \citet{perley:2009}. 
The kinetic energies injected in the spike 
and extended emission   
components are $E_{\rm kin}^{\rm spike} = 7 \ 10^{50}$ erg  and 
$E_{\rm kin}^{\rm ee} = 50 \ E_{\rm kin}^{\rm spike}$. 
We adopt $\epsilon_e = \epsilon_B^{1/2} = 5 \ 10^{-2}$, $p = 2.5$ in the shocked external medium together with
$n=10^{-3}$ cm$^{-1}$ and $z=0.5$. 
The steeper thin lines at late
times correspond to a conical jet (seen on-axis) of opening angle $\theta_{\rm jet}=0.06$ rad.}
\label{fig_scen1_fs}
\end{figure*}

To obtain a rebrightning at one day we adopted $\Gamma_{\rm ee}=20$ and $\Gamma_{\rm spike}=300$. The 
outflow lasts for a total duration of 100 s (1s for the 
the spike and 99 s for the tail). We injected a kinetic energy $E_{\rm kin}=7\,10^{50}$ erg in the spike
and 50 times more in the tail.
It can be seen that the results, shown in \fig{fig_scen1_fs}, are consistent with the available data and upper
limits except possibly after the peak of the rebrightening where the decline of the synthetic light curve
is not steep enough. This can be corrected if a jet break occurs close to the peak, which is possible if
the jet opening angle $\theta_{\rm jet}$ is on the order of $1/\Gamma_{\rm ee} \simeq 0.05$ rad.  
This beaming angle is somewhat smaller than the values usually inferred from observations of short burst afterglows 
(see e.g. \citealt{burrows:2006, grupe:2006}) or suggested by simulations of compact binary mergers (see e.g. \citealt{rosswog:2002}). 

An example of a light curve with a jet break 
is shown in \fig{fig_scen1_fs}, assuming that the jet has an opening angle of 3.4$^\circ$ (0.06 rad) and is seen on-axis.
A detailed study of the jet-break properties is beyond the scope of this paper and we therefore did not consider 
the case of an off-axis observer and neglected the lateral spreading of the jet, expected to become important
when $\Gamma\lsim 1/\theta_{\rm jet}$. Detailed hydrodynamical studies (see e.g. \citealt{granot:2007, 
zhang:2009, vaneerten:2011, lyutikov:2011}) tend to show, however, that as long as
the outflow remains relativistic, the jet-break is more caused by the ``missing'' sideways emitting material 
than by jet angular spreading. 

We finally checked how our results are affected if different model parameters are adopted. If the density $n$ of the 
external medium is increased or decreased, similar light curves can be obtained by changing the Lorentz factors
(to still achieve the rebrightning at one day) and the microphysics parameters (to recover the observed flux). For example,
increasing the density to $n=0.1$ cm$^{-3}$ requires $\Gamma_{\rm ee}\simeq 10$ (keeping $\Gamma_{\rm spike}=300$) and 
$\epsilon_e=\epsilon_B^{1/2}=0.02$. Conversely, with $n=10^{-5}$ cm$^{-3}$, $\Gamma_{\rm ee}\simeq 35$ and 
$\epsilon_e=\epsilon_B^{1/2}=0.08$ are needed. 

If the kinetic energy of the outflow is increased (resp. decreased) because the radiative efficiency $f_{\rm rad}$ is
lower (resp. higher) or the redshift higher (resp. lower), light curves agreeing with the data can again be obtained 
by increasing (resp. decreasing) $\Gamma_{\rm ee}$ and decreasing (resp. increasing) $\epsilon_e=\epsilon_B^{1/2}$. Also 
note that the spread of the Lorentz factor $\delta \Gamma_{\rm ee}$ around $\Gamma_{\rm ee}$ at the end of the 
prompt phase has to be limited to ensure that
the slower material is able to catch up 
in a sufficiently short time to produce an effective rebrightening. 
In the case shown in \fig{fig_scen1_fs} we have $\delta \Gamma_{\rm ee}/\Gamma_{\rm ee}=0$ but we have checked from the numerical simulation that 
acceptable solutions can be 
obtained as long as $\delta \Gamma_{\rm ee}/\Gamma_{\rm ee}\lsim 0.2$. This configuration is for example naturally
expected after an internal shock phase where fast and slow parts of the flow collide, resulting in a shocked region with
a nearly uniform Lorentz factor distribution (see e.g. \citealt{daigne:2000}). 

\subsubsection{Long-lived reverse shock model}
\label{sect_refreshed_rs}

\begin{figure*}
\begin{tabular}{cc}
\includegraphics[width=0.45\textwidth]{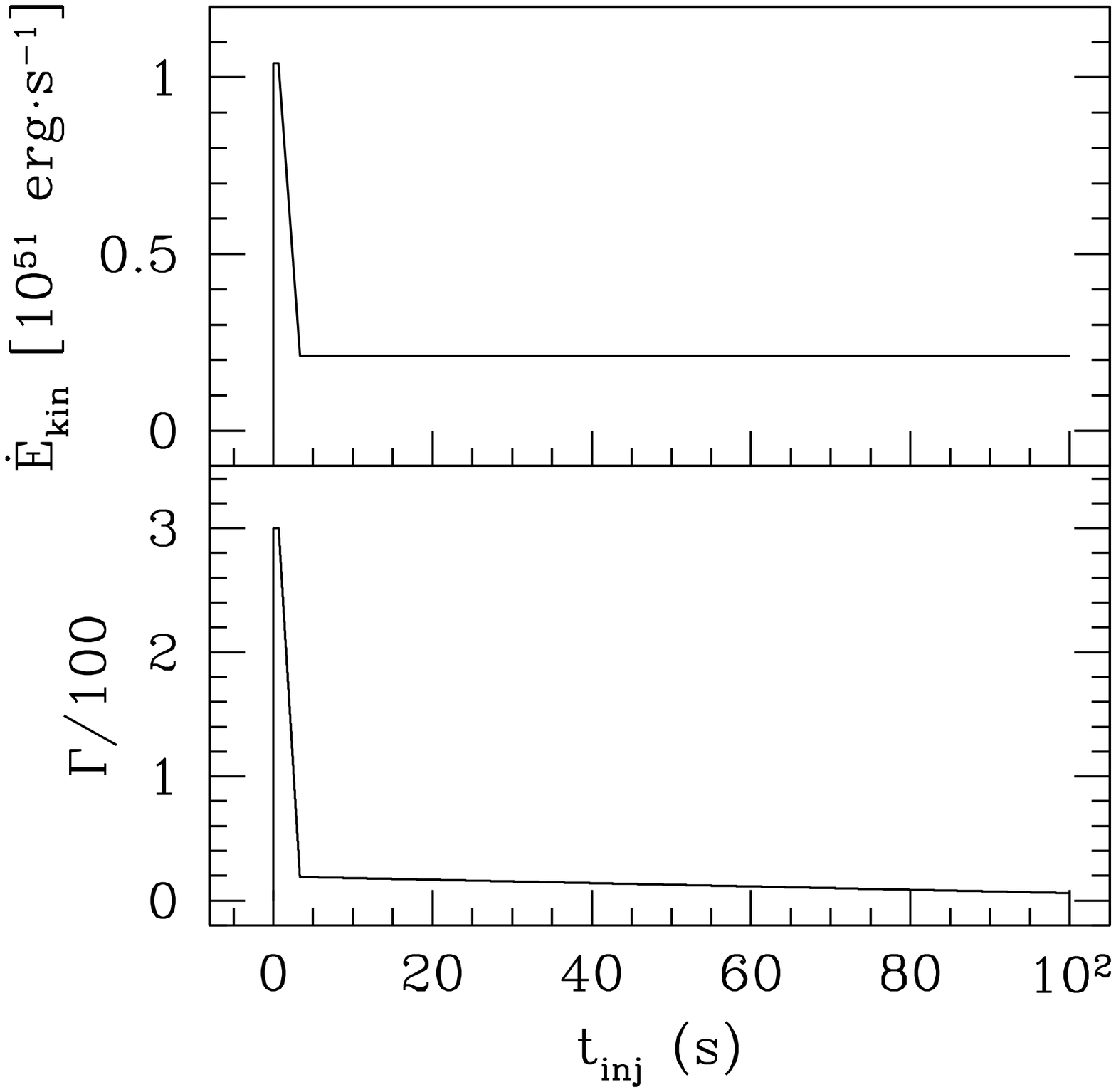} & \includegraphics[width=0.45\textwidth]{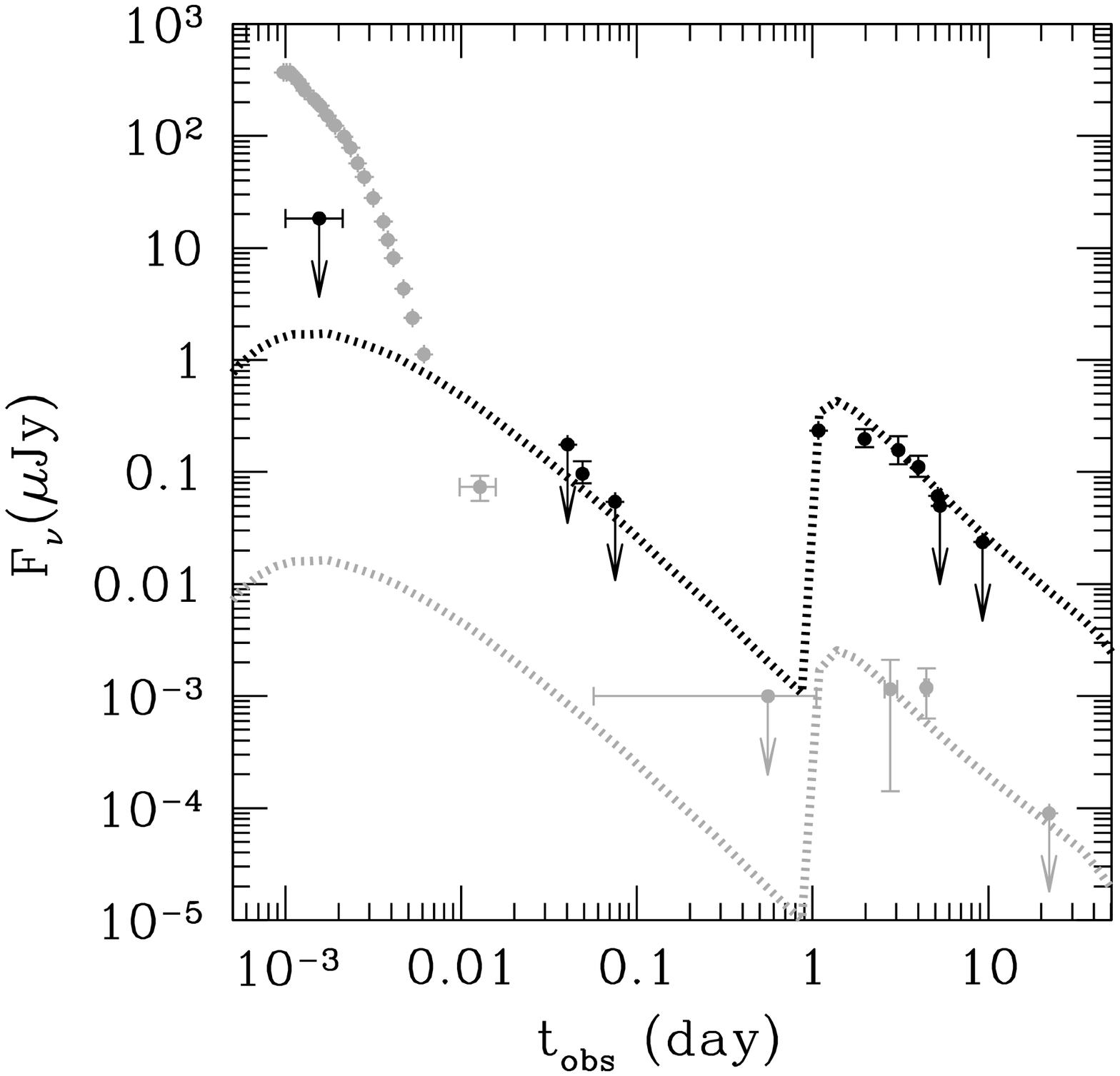} 
\end{tabular}
\caption{\textbf{Refreshed shocks: long-lived reverse shock model.} 
\textit{Left panel}: 
initial distribution of the Lorentz factor (lower part) and kinetic power (upper part) in the flow as a function 
of injection time $t_{\rm inj}$. 
\textit{Right panel}: 
synthetic light-curves at $2$ eV (black, dotted line) and $10$ keV
(grey, dotted line) together with the data.
The kinetic energies in the spike 
and extended emission  
components are $E_{\rm kin}^{\rm spike} = 7 \ 10^{50}$ erg  
and $E_{\rm kin}^{\rm ee} = 30 \ E_{\rm kin}^{\rm spike}$. 
The density of the external medium,
redshift and slope $p$ of the electron distribution are the same as in \fig{fig_scen1_fs}. 
The adopted microphysics parameters are
$\epsilon_e = \epsilon_B^{1/2} = 0.16$ in the shocked ejecta.}
\label{fig_scen1_rs}
\end{figure*}

If the afterglow is produced by the reverse shock, similar good fits of the data can be obtained. 
\fig{fig_scen1_rs} shows an example
of synthetic light curves for $E_{\rm kin}^{\rm spike} = 7 \ 10^{50}$ erg  and $E_{\rm kin}^{\rm ee} = 30 \ E_{\rm kin}^{\rm spike}$,
$\epsilon_e = \epsilon_B^{1/2} = 0.16$, $p = 2.5$ in the shocked ejecta and
$n=10^{-3}$ cm$^{-3}$. The microphysics parameters have to be higher than in the forward shock
case because the reverse shock is dynamically less efficient, which requires a higher radiative efficiency to obtain the same 
observed fluxes. The Lorentz factor distribution in the ejecta is also slightly different to guarantee that the light
curve {\it (i)} goes through the optical point at 0.05 day and {\it (ii)} decays steeply after the peak. No jet break has to be invoked 
here because the decay rate (in contrast to what happens in the forward shock model) depends on the distribution of energy 
as a function of the Lorentz factor in the ejecta.

Again, if the external density and kinetic energy of the flow are varied, satisfactory fits of the data can
be recovered by slightly adjusting the Lorentz factor and microphysics parameters.  

\subsection{Density clump in the external medium}

\begin{figure*}
\begin{tabular}{ccc}
\includegraphics[width=0.3\textwidth]{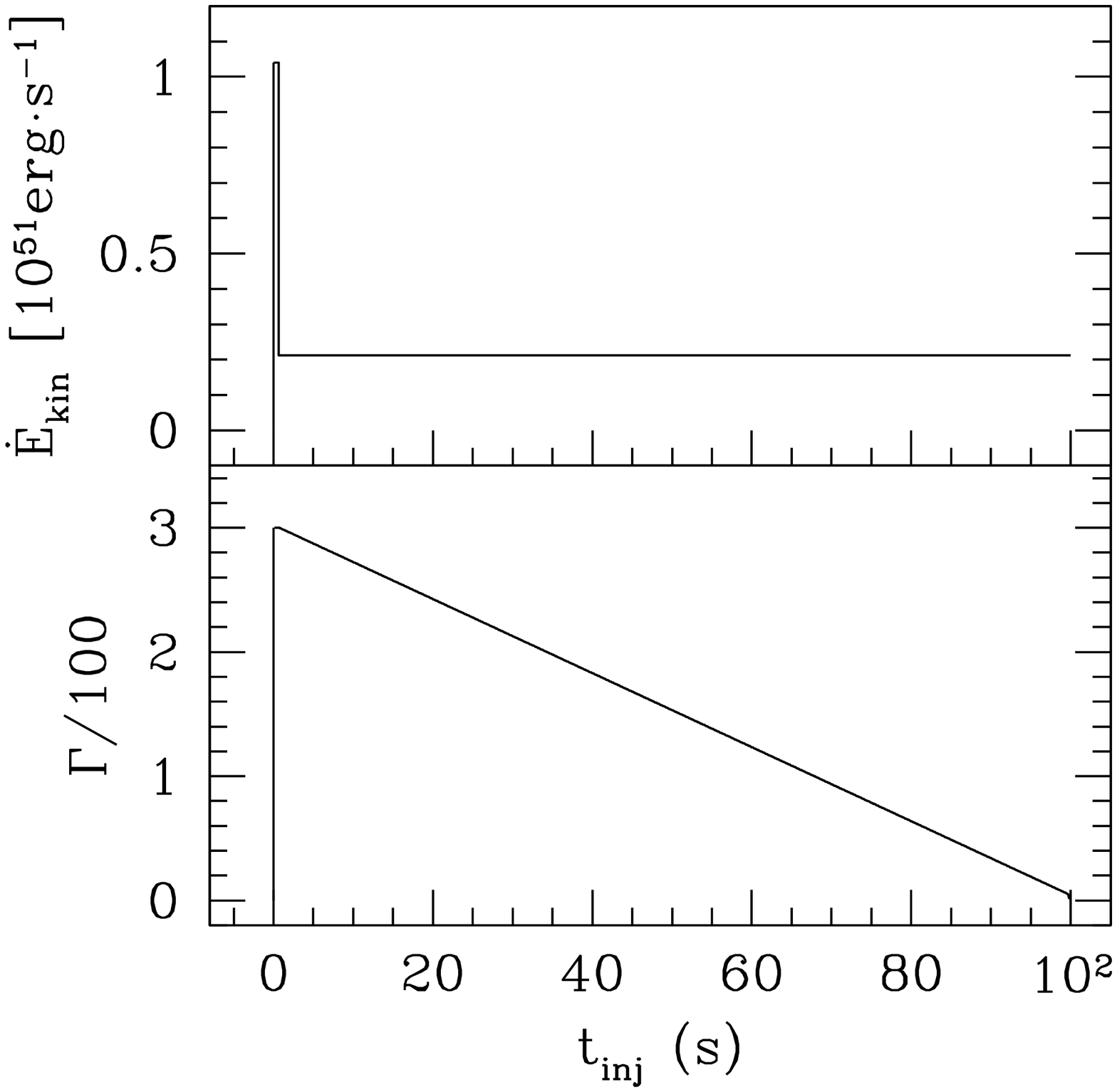} & \includegraphics[width=0.3\textwidth]{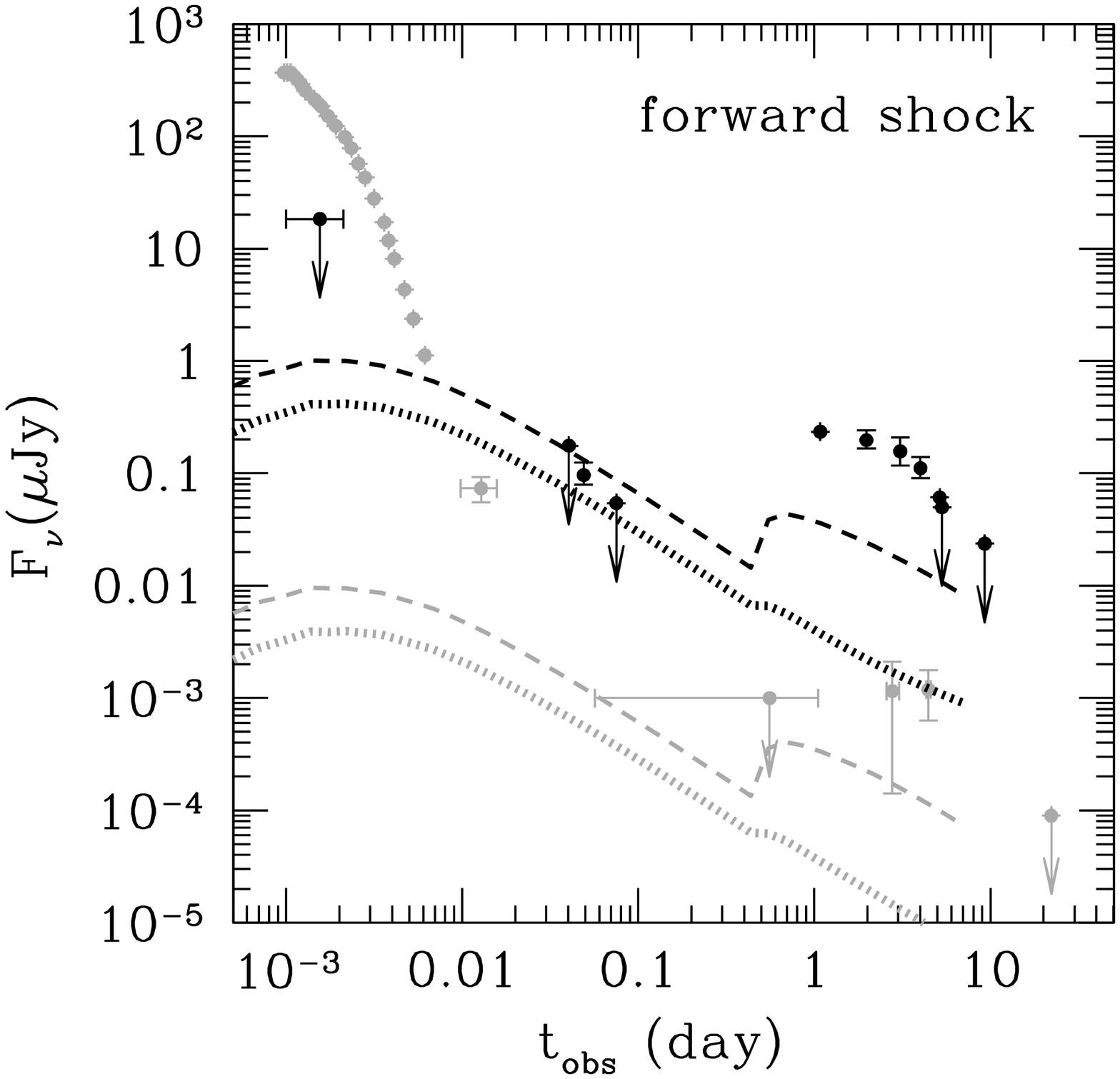} & \includegraphics[width=0.3\textwidth]{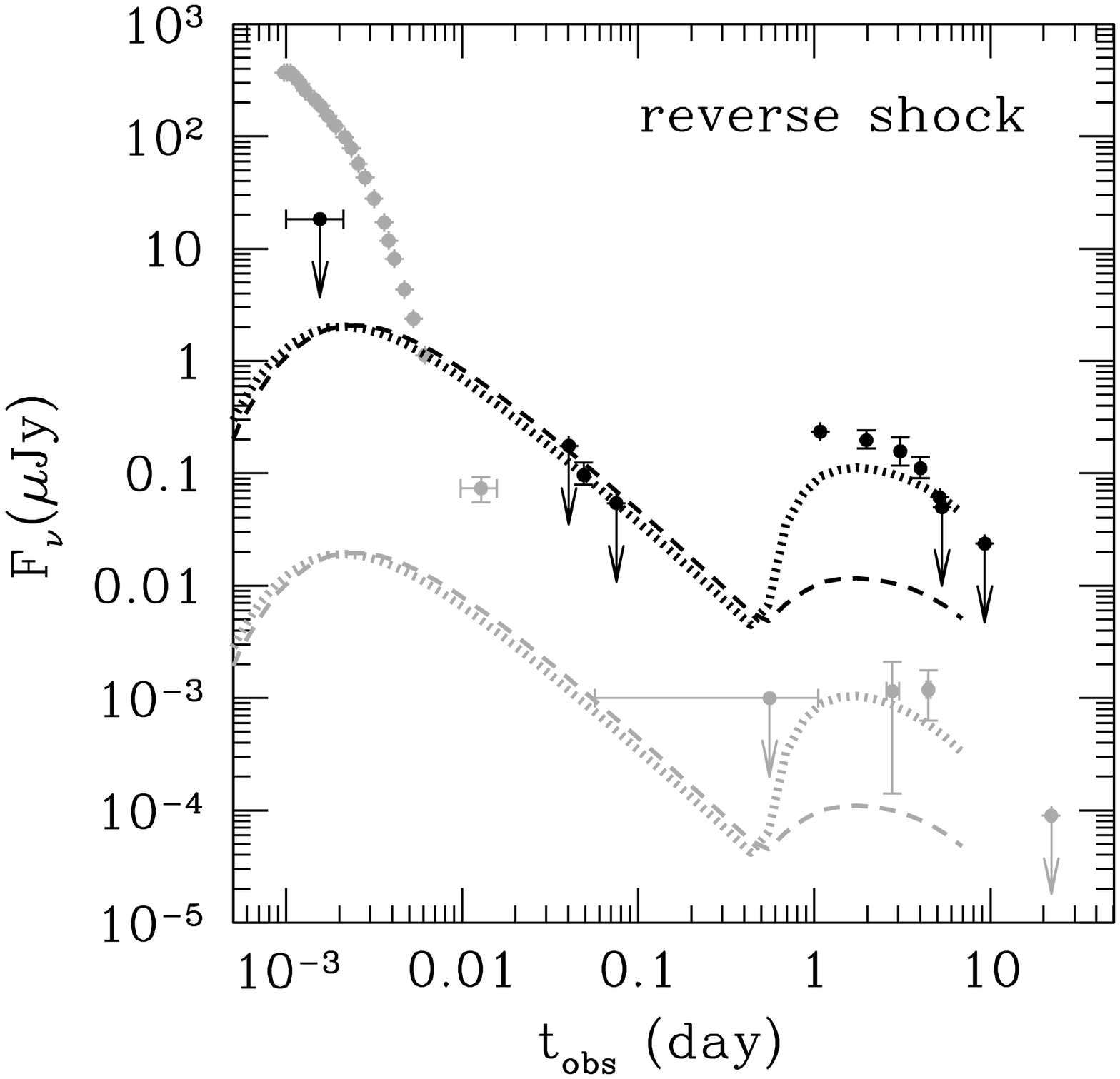} 
\end{tabular}
\caption{\textbf{Density clump: forward and long-lived reverse shock models.}
\textit{Left panel}: 
initial distribution of the Lorentz factor (lower part) and kinetic power (upper part) in the flow as a function of injection time $t_{\rm inj}$.
\textit{Middle and right panels}: 
synthetic light-curves at 2 eV (black) and 10 keV (grey) for the forward and reverse 
shocks, using either the simple one-zone (dotted lines) or the detailed multi-zone
(dashed lines) model. 
The kinetic energies in the spike 
and extended emission
components 
are $E_{\rm kin}^{\rm spike} = 7 \ 10^{50}$ erg  and $E_{\rm kin}^{\rm ee} = 30 \ E_{\rm kin}^{\rm spike}$. 
The adopted microphysics parameters
are $\epsilon_e = \epsilon_B^{1/2} = 10^{-2}$ (forward shock, middle panel) and  
$\epsilon_e = \epsilon_B^{1/2} = 0.07$ (reverse shock, right panel).   
See text for the prescription adopted for the density clump.}
\label{fig_density_bump}
\end{figure*}

We now investigate the possibility that the rebrightening is caused by the encounter of the decelerating ejecta with
a density clump in the external medium. For illustration, we adopted a simple distribution of the Lorentz factor that linearly decreases with
injection time from 300 to 2 so that, in the absence of the density clump, afterglow light curves from either 
the forward or reverse shocks would be smooth and regular (we have checked that the exact shape of the low Lorentz factor tail is not crucial in this scenario). To model the clump, we assumed that the circumburst medium
is uniform (with $n=10^{-3}$ cm$^{-3}$) up to 1.7 $10^{18}$ cm (0.55 pc) and that the density then rises linearly
to $n=1$ cm$^{-3}$ over a distance of $10^{18}$ cm (0.32 pc). The ejecta is strongly decelerated after entering the high-density region
and we find that the forward shock is still inside the clump at the end of the calculation (at $t_{\rm obs}=8$ days).

\subsubsection{Forward shock model}
As \citet{nakar:2007} showed by coupling their hydrodynamical calculation to a detailed radiative code,
a density clump in the external medium has little effect on the forward shock emission. 
Therefore a clump cannot produce the rebrightening in GRB 080503. 
In the simple case where the shocked medium is represented by a single zone, the effect of the clump is barely visible.
With the detailed multi-zone model a stronger rebrightening is found, because the effects of the 
compression resulting from 
the deceleration of the flow are better described,
but even in this case the calculated flux remains nearly one order of magnitude below the data.    

\fig{fig_density_bump} illustrates these results and confirms that the forward shock emission
does not strongly react to the density clump. Even if, from an hydrodynamical point of view, the forward shock is 
sensitive to the clump, the observed synchrotron emission is only moderately affected because 
the increase in upstream density is nearly counterbalanced by the decrease of Lorentz
factor in the shocked material.
Of course, spectral effects complicate the picture, but the essence of the result remains the same (see \citealt{nakar:2007} for details).

\subsubsection{Possible evolution of the microphysics parameters}
\label{paragraph_evol_micro}

\begin{figure}
\begin{tabular}{c}
\includegraphics[width=0.45\textwidth]{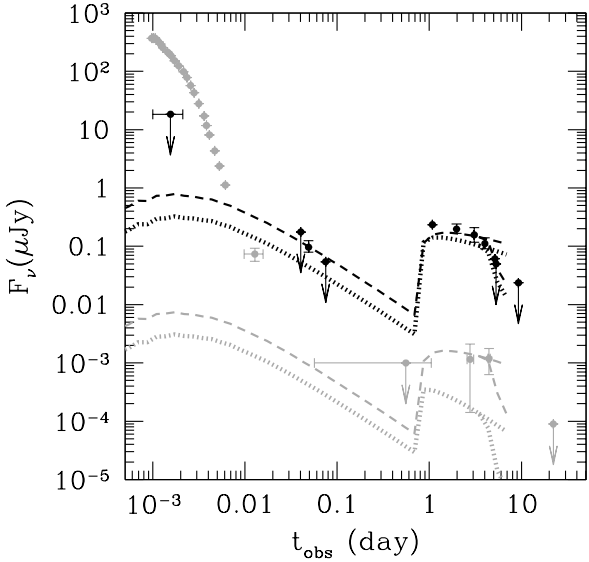} 
\end{tabular}
\caption{\textbf{Density clump: forward shock model with varying microphysics parameters.}
Synthetic light-curves at 2 eV (black) and 10 keV (grey) when the microphysics parameters of the forward shock
are changed at the density clump: $\epsilon_e$ is 
increased by a factor $5$ from an initial value of $10^{-2}$, keeping the prescription
$\epsilon_e = \epsilon_B^{1/2}$. The Lorentz factor, injected kinetic power in the outflow, 
and the density distribution in the external medium are the same as in \fig{fig_density_bump}. 
The dotted and dashed lines 
correspond to the single and multi-zone models for the shocked region. 
The same simulations with the addition of a jet-break 
(assuming $\theta_\mathrm{jet} = 0.08$ rd) are also shown.  }
\label{fig_micro_var}
\end{figure}

In view of the many uncertainties in the physics of collisionless shocks it is often assumed for simplicity,
as we did so far,  
that the microphysics
redistribution parameters $\epsilon_e$ and $\epsilon_B$ stay constant during the whole afterglow evolution.
However, particle-in-cell simulations of acceleration in collisionless shocks (see e.g. \citealt{sironi:2011}) do not show any evidence of universal values of the parameters. 
If $\epsilon_e$ or/and $\epsilon_B$ are allowed to change during afterglow
evolution, the problem of the forward shock encountering a density clump can be reconsidered, now with the possibility
of a sudden increase of radiative efficiency triggered by the jump in external density. 

\fig{fig_micro_var} shows the resulting light curves when the microphysics parameters $\epsilon_e$ and $\epsilon_B$ 
of the forward shock are increased at the density
clump. If the prescription $\epsilon_e= \epsilon_B^{1/2}$ is maintained, no satisfactory solution can be found 
in the simple model where the shocked medium is represented by one single zone. In this case, 
the optical frequency lies between the injection and cooling frequencies ($\nu_i<\nu_{\rm opt}<\nu_c$)
while the X-ray frequency satisfies $\nu_{\rm X}>\nu_c$ so that 
the visible and X-ray flux densities depend on the microphysics parameters in the following way \citep{panaitescu:2000}
\begin{equation}
f_{\nu,{\rm opt}}\propto \epsilon_e^{p-1}\epsilon_B^{p+1\over 4}   \  {\rm and} \ \  
f_{\nu,{\rm X}}\propto \epsilon_e^{p-1}\epsilon_B^{p-2\over 4} .
\end{equation} 
With the prescription $\epsilon_e=\epsilon_B^{1/2}$ we obtain 
$f_{\nu,{\rm opt}}\propto \epsilon_e^{3p-1\over 2}$ and $f_{\nu,{\rm X}}\propto \epsilon_e^{3 p-4 \over2}$
The optical flux is therefore much more sensitive than the X-ray flux to a change of the
microphysics parameters and a simultaneous fit of the data in both energy bands is not possible.
A simple solution to this problem is to change $\epsilon_e$ alone, keeping
$\epsilon_B$ constant. In this case, increasing $\epsilon_e$ by a factor of 25 (from 0.01 to 0.25) is required
to reproduce the rebrightening in both the X-ray and visible ranges.

In the more detailed model with a multi-zone shocked region, the situation is
different. In the shells that contribute most to the emission, we find that both $\nu_{\rm opt}$ and $\nu_{\rm X}$
are larger than $\nu_c$ and therefore $f_{\nu,{\rm opt}}$ and $f_{\nu,{\rm X}}$ depend in the same way 
on the microphysics parameters. It is then possible to achieve a satisfactory solution 
(dashed lines in \fig{fig_micro_var}) that keeps the prescription $\epsilon_e=\epsilon_B^{1/2}$ 
(with $\epsilon_e$ increased by a factor $5$). As in \sect{sect_refreshed_fs} we introduce a jet break
(now assuming $\theta_{\rm jet} = 0.08$ rd) to account for the decay of the optical
flux following the peak of the rebrightening. Notice that the decay is steeper here (compare 
\fig{fig_scen1_fs} and \fig{fig_micro_var}) owing to the rapid decrease of the Lorentz factor inside the clump. 

\subsubsection{Long-lived reverse shock model}
With the simple one-zone model, the reverse shock emission is found to be much more sensitive 
to the density clump than the forward shock emission. Indeed, when the ejecta starts to be decelerated, 
its bulk Lorentz factor suddenly decreases and slow shells from the tail material pile up at a high 
temporal rate and with a strong contrast 
in Lorentz factor. 
These two combined effects lead to a sharp rise of the flux from the reverse shock. Synthetic light curves 
showing a satisfactory agreement with the data are shown in \fig{fig_density_bump}. 

However, the detailed multi-zone model gives different results, where the rebrightening is dimmer 
and cannot fit the data. The main reason is that the higher contrast in Lorentz factor,
leading to a higher specific dissipated energy, now  
only concerns the freshly shocked shells, while in the single zone model it is applied to the whole shocked 
region. 
This example (as well as the one already discussed in \sect{paragraph_evol_micro}) shows that using a detailed description of the
shocked material can be crucial 
when dealing with complex scenarios (i.e. not the standard picture where the blast-wave propagates 
in a smooth external medium, with constant microphysics parameters)\footnote{In the refreshed-shock
scenario (\sect{refreshed_shocks}) where the dynamics is simpler, the single and multi-zone models give comparable results.}. 

It is still possible to fit the data by increasing the microphysics parameters during the propagation in the clump. 
However, this seems less natural than for the forward shock (\sect{paragraph_evol_micro}) because 
the upstream density of the reverse shock does not change.
On the other hand, a modification of the microphysics parameters could still be due to the sudden increase in 
the reverse shock Lorentz factor triggered by the clump encounter.

\section{Conclusion}
\label{sect_conclusions}

GRB 080503 belongs to the special group of short bursts where an initial bright spike is followed 
by an extended soft emission of much longer duration.
It did not show a transition to a standard afterglow after the steep decay observed in
X-rays at the end of the extended emission. This behavior has been observed previously in 
short bursts, but GRB 080503 was peculiar because it exhibited a spectacular rebrightening 
after one day, both in X-rays and the visible. The presence of the extended emission prevents one
from classifying GRB 080503 on the basis of duration only, but the lack 
of any candidate host galaxy at the location of the burst and the vanishing spectral lag of 
the spike component are consistent with its identification as a type-I event resulting from 
the coalescence of a binary system consisting of two compact objects. 

From its formation to the coalescence, the system can migrate to the external regions of the 
host galaxy allowing the burst to occur in a very low density environment, accounting for
the initial lack of a detectable afterglow. To explain the late rebrightening, we
considered two possible scenarios -- refreshed shocks from a late supply of energy or a density clump
in the circumburst medium -- and two models for the origin of the afterglow, the standard one where 
it comes from the forward shock and the alternative one where it is made by a long-lived reverse shock.   

In the refreshed-shock scenario we supposed that the initial spike was produced by fast 
moving material (we adopted $\Gamma=300$) while the one making the soft tail was slower ($\Gamma\sim 20$). 
Initially, only the spike material is decelerated and contributes to the afterglow until   
the tail material is eventually able to catch up, which produces the rebrightening. 
Both the forward and reverse shock models provide satisfactory fits of the data under the 
condition that the material making the tail has a limited spread in Lorentz factor
$\delta \Gamma/\Gamma\lsim 0.2$. This allows the rise time of the rebrightening to be sufficiently short.
This condition might be satisfied from the beginning but can also result from a previous 
sequence of internal shocks that has smoothed most of the fluctuations of the Lorentz factor initially
present in the flow. In addition, a jet break is required in the forward shock case
to reproduce the steep decline that follows the rebrightening. This implies that the jet should be beamed within
an opening angle of 3 - 5$^\circ$, which appears somewhat smaller than the values usually preferred for type-I bursts. 
In the long-lived reverse shock model a jet break is not necessary 
because the shape of the light curve now depends on the energy distribution in the ejecta, which can be
adjusted to fit the data. 

In the scenario where a density clump is present in the
burst environment, the rebrightening resulting from the forward shock 
is weak, in agreement with
the previous work of \citet{nakar:2007}. We performed the calculation in 
two ways: first with a simple method where the shocked material was represented
by one single zone, then using a more detailed, multi-zone approach. The impact of the
clump was barely visible in the first case. The rebrightening was larger 
in the second one but still remained nearly one order of magnitude below the data. 
We then considered the possibility that the shock microphysics might change inside the clump.
We found that by increasing $\epsilon_e$ by a factor of five (and with the prescription that
$\epsilon_e=\epsilon_B^{1/2}$) 
it was possible to fit the data with the multi-zone model under the additional condition to have
a jet break at about 2 - 3 days (corresponding to a jet opening angle $\lsim 5^{\circ}$).
With the simplified
model the results were more extreme, imposing to increase $\epsilon_e$ alone by a very large factor of 
25. 

If the afterglow is made by the reverse shock, the effect of the clump is strong with the simple model. 
It is however much reduced with the detailed model and the
observed rebrightening cannot be reproduced, the synthetic light curve lying nearly one order of magnitude 
below the observed one. 
It appears that only the multi-zone approach provides 
a proper description of the compression resulting from the encounter with the density barrier. Conversely,
in the refreshed-shock scenario the simple
and detailed models give comparable results. 

From the different possibilities we considered, which could explain the late rebrightening in GRB 080503, 
several appear compatible with the data, but none is clearly favored. The refreshed-shock scenario may seem more natural
because the initial spike and extended emission probably correspond to different phases of central engine activity.
It is not unreasonable to suppose that the material responsible for the extended emission had a lower Lorentz factor,
as required by the refreshed-shock scenario. Then, both the forward and reverse shock models lead to satisfactory fits
of the X-ray and visible light curves, if two conditions on the Lorentz 
factor distribution and jet opening angle (see above) are satisfied.
The density clump scenario does not seem able to account for the rebrightening if the afterglow is made
by the reverse shock. The conclusion is the same with the forward shock, except if the microphysics
parameters are allowed to change when the shock enters the clump.

\begin{acknowledgements} 
The authors acknowledge the French Space Agency (CNES) for financial support.
R.H.'s PhD work is funded by a Fondation CFM-JP Aguilar grant.
\end{acknowledgements}

\bibliographystyle{aa} 
\bibliography{main}

\end{document}